# Harnessing AI in Secondary Education to Enhance Writing Competence

Eyvind Elstad[1] (University of Oslo) & Harald Eriksen[2] (Oslo Metropolitan University)

**Abstract**

The emergence of free generative artificial intelligence (AI) tools such as ChatGPT offers significant implications for developing writing skills in secondary education. Writing competence provides the technical foundation for text creation, while writing voice adds unique personal expression. These facets contribute to literacy, which, along with reading competence, plays a central role in education.  This study explores AI's potential effects on students' writing competence and the cultivation of a personal writing voice, emphasizing the dichotomy between technological augmentation and potential risks of dependency and plagiarism. We explore the pros and cons of integrating AI into the writing process, considering insights from recent investigations that highlight diverse student perspectives on AI's role. Our discussion underscores the importance of process-based assessments, task designs fostering creativity and personal reflection, and AI as a supplementary educational tool rather than a replacement for teacher guidance. We also address critical questions about AI's role in various stages of the writing process, examining pre-writing, writing, and revision phases, and the feedback mechanism. The paper discusses the importance of process-based assessments, the design of tasks that encourage deep personal reflection and creativity, and the role of AI as an educational tool that supplements rather than replaces the teacher's guidance. Ultimately, the pragmatic use of AI in education necessitates a balanced approach, which might involve technological controls, innovative assignments, and an emphasis on critical thinking and personal judgment to maintain writing as a deeply human and expressive endeavor. We argue for a balanced approach to AI utilization in education, stressing the need for technological controls, innovative assignments, and an emphasis on critical thinking to preserve the deeply human and expressive nature of writing.

**Introduction**

From 30 November 2022, a free version of generative artificial intelligence (AI) was made available, which sent shockwaves through educational institutions. Suddenly, AI could solve typical assignments in schools, and the production of academic text became simple. What we do not know yet is how the availability of generative AI affects the development of students' writing competency (Baron, 2023). Writing ability refers to the overall skill and proficiency in writing (Bereiter & Scardamalia, 2013) while writing voice refers to the characteristic expression of a given writer (Engebretsen, 2021). It encompasses how the writer's personality, tone, and individual style come through in their writing. Both are essential to effective writing. While writing ability provides a technical foundation, writing voice adds the personal touches that make writing distinctive and engaging. Together, they contribute to a writer's overall effectiveness and impact. In this article, we focus attention on secondary education and investigate two questions: How can AI be a partner in the development of writing competence (Baron, 2023)? And how can schools prevent students from using AI in ways that weaken their development of writing competence and learning (Hodges & Kirschner, 2024)?

---

[1] Email: eyvind.elstad@ils.uio.no
[2] Email: haralde@oslomet.no

Mastering a language means being able to create expressions and understand what other people say. The ability to use language is innate and marked by creativity. For example, small children might say 'mouses' instead of 'mice', a non-intuitive form that they may not have previously heard. This illustrates the brain's ability to produce logical grammatical inferences (Monti et al., 2009). Although the ability to learn language is innate, language learning occurs through environmental influence; we do not develop language in isolation. At the same time, language skills contribute to improving our ability to think. To write is to think, and when we think, we can learn (MacArthur & Graham, 2016; Willingham, 2017). Children naturally learn the oral version of their language arts by listening and observing, but learning to read and write is not innate and is much more difficult than learning to speak (Geary, 2007). Writing is so difficult that a student's own motivation is not sufficient to achieve that goal. Therefore, instruction in writing is needed (Applebee & Langer, 2015).

Helping students find their own writing voices is an important part of writing instruction. A writing voice is what gives a text life and makes it an authentic expression of the author's views (Sperling & Appleman, 2011). However, a certain level of writing competence is required to be able to experiment and express oneself in a distinctive style (Pinker, 2015). Writing competence is about the ability to create a well-composed text (Bereiter & Scardamalia, 2013). The development of a personal writing voice and the cultivation of writing competence often go hand in hand and can strengthen each other. We base these considerations on a functional view of students' writing (Jones, 2004); we see writing as a meaningful and practical skill used in different contexts and for different purposes. Over time, students should develop into independent and proactive individuals who can take care of their lives in a society characterized by literacy (Biesta & Stengel, 2016).

It takes several years to develop the ability to become a functional reader and writer, and that training takes place throughout the entire school journey (Bereiter & Scardamalia, 2013). Together, writing competence and reading competence make up the concept of literacy (Gee, 2012); both are important in the teaching of many school subjects, but especially in teaching language arts. Reading and writing are in many ways two sides of the same coin: by reading, the student is exposed to a wealth of words, which can provide a more detailed understanding of how words are used (Gregg & Steinberg, 2016). The student also learns how texts are organized and becomes acquainted with various genres, grammatical structures, and sentence construction, which is useful for creating their own texts. Moreover, texts that the student reads can provide ideas for various ways to argue. Reading is a nutrient for the brain, while the writing process is exercise for the brain (Willingham, 2017).

We lack systematic measurements of the development of writing skills in schools, partly because it is challenging to develop standardized assessments that capture enough aspects of writing competence. However, tests in the large-scale international reading surveys (Programme for International Student Assessment, PISA) and progress in International Reading Literacy Study, PIRLS) may provide an indicator of at least some of the skills relevant to writing, such as text comprehension and vocabulary (OECD, 2023). Many find it worrying that a significantly larger proportion of language arts students in the Nordic countries are performing at a low level in reading than was previously the case. PIRLS results showed that Nordic 10-year-olds in 2021 read significantly worse than their counterparts in 2016; the percentage of students at or below a low level of proficiency in reading increased in that period (Komakhidze et al., 2024). The PISA survey also shows an increase in weak readers. When reading skills weaken, it can have a negative impact on writing skills. The term 'functional illiterates' is used about people – who despite having received basic schooling and able to recognize words and letters – still have difficulties applying reading and writing skills at the level necessary to function effectively in daily life, including the workplace (Vágvölgyi et al., 2016). Students performing at a low level in reading are at risk of also developing into functional illiterates in writing. The school is an indispensable arena for bolstering these students' writing competence.

What does the availability of AI mean for teaching the language arts? A survey conducted by Bueie (2024) indicates an increase in Norwegian secondary students' use and perception of chatbots such as ChatGPT for writing. At the same time, students appear to have developed a better understanding of technology and are also more conscious of how they want to use it. Many students consider the use of chatbots to be cheating and would not want to delegate their writing work to chatbots. Their use of it as a tool in the writing process is, however, more acceptable. The school thus plays a role in students' understanding and use of generative AI. When students are allowed to use chatbots in teaching, they may obtain a better understanding of the tools than if they only used them on their own. At present, we do not know well how advantageous AI tools are for developing writing competence (Roe et al., 2023). Therefore, we must largely rely on conclusions based on logic and reason.

**Writing and AI**

Writing in schools is an important context for writing instruction (Applebee & Langer, 2015), and AI can be a potential resource in efforts to improve writing skills (Kuhail et al., 2023; Bibi & Atta, 2024; Marzuki et al., 2024), but it can also lead to students failing to engage in the academic work required to develop a genuine writing voice (Pinker, 2015). This duality in the view of writing instruction is already embedded in the framework of school activities. On one hand, national examination regulations stipulate in several countries that students are not allowed to use generative AI in connection with examinations. Thus, students have a clear incentive to write texts without accessing AI tools. This is an important clarification.

On the other hand, generative AI, like other online resources, has the potential to be a resource in mastering the curriculum and developing academic skills, including writing (Fadli et al., 2024; Fitria, 2023). The critical question is how students can best use such resources and avoid weakening the development of their writing competence by copying text created by AI (Tlili et al., 2023; Hodges & Kirschner, 2024). This raises important questions about the teacher's role in writing instruction (Jeon & Lee, 2024).

The functional perspective on the development of a writing voice emphasizes that writing has applications in countless contexts (Vágvölgyi et al., 2016). In this article, we argue that students in study preparatory programs in secondary schooling should learn to use generative AI in ways that promote learning, including developing a language style that is also functional in higher education. Similarly, we argue that learning environments must be designed so that students avoid taking learning-inhibiting shortcuts when working on their writing (Hodges & Kirschner, 2024). For example, there are software programs (for instance, Safe Exam Browser) that prevent students from leaving the site before submitting their written work to the teacher (Cronqvist & Kortesaari, 2023; Frankl et al., 2017).

4Whether AI can function as an effective tool for students in their writing process is controversial, and there are yet no solid studies clarifying this issue. Some teachers' experiences are disappointing (Livingstone, 2024), but others argue that students can use AI as a starting point for their writing; the text produced can then be edited and improved by students adding their own voice through choice of words, logical argumentation, rhetorical techniques, and fact sharing (Kuhnke et al., 2023). In this case, however, the use of AI can be compared to the use of anabolic steroids in sports: both can potentially improve a person's performance and may yield more impressive outcomes. But both AI in education and steroids in sport raises questions about cheating and whether the results truly reflect an individual's own abilities and effort (Hogdes & Kirschner, 2024). The rest of this article is related to students' struggle with writing in secondary school.

**Writing, Reading, and Thinking**

Learning that promotes understanding is based on thinking (Willingham, 2017). Through the writing process, insight can be created, and therefore the material on which writing is based is often influenced by what the writer has read (Genlott & Grönlund, 2013). Hence, we must view writing competence in relation to the ability to understand and interpret written material through reading. Written materials here can broadly be interpreted as texts, images, videos, and audio formats like music, but the focus in this article is on texts comprised of words. Texts produced by generative AI can serve as the basis for writing, but the question is how beneficial they are for the learning process. A person's knowledge base is clearly a factor that matters for the ability to express oneself through a written text, in addition to grammatical and orthographic competence (MacArthur & Graham, 2016). The writing process often finds itself at the intersection between technical skill and personal expression. Writing competence provides the student with the tools, while the writing voice gives the texts a personal character.

The distinction between the material on which writing is based and the learning individual's own expression does not have to be entirely sharp; we can sometimes speak of hybridization in the text creation process, resulting in a text with mixed origins (Baron, 2023). One example is text excerpts that make up a digital scrapbook, such as material that students copy from digital learning materials into a file repertoire or compilation file that forms a textual base for their own writing. Texts produced by generative AI and snippets of documents retrieved from the internet could also be parts of the text base in a digital scrapbook.

If a student reads the text repertoire to prepare for tests and oral discussions, the texts that AI produces could function as one possible source among sources, as long as the student is aware that AI can hallucinate and produce incorrect information (Zhou et al., 2024). AI can also be a useful tool for clarifying the meaning of words and phrases that students encounter in texts. In the student's written work, the use of a digital scrapbook will require citations or students to imprint their own mark on the text. If a student copies text excerpts unaltered from a digital scrapbook and submits them as their own text without citing a source, that is plagiarism (Hutson, 2024). The range from lawful use of AI in text creation (i.e., with source use declared) to plagiarism encompasses many intermediate positions: 'is there ever enough personal treatment and rewriting of a text to justify omitting the source? We need clear distinctions in this continuum between plagiarism and acceptable use of AI, but national regulations are thus far unclear about where this intermediate zone may lie.

One might wonder about just how dangerous it is to use AI in text production (Baron, 2023). When can we say that AI is a partner in our writing process, and when can we say that AI has created a text? Baron (2023, p. 1979) uses collaboration as a theoretical lens for writing with the aid of AI and argues that we must include the human in this picture. In other words, Baron believes that if AI is to be used, the student's collaboration must ensure authenticity.

Thinking-writing is intended to help an author organize subject material or generate ideas that can be part of a larger writing process, which we describe more below. In such cases, the learning individual primarily activates thoughts for themselves (Dysthe et al., 2015). The teacher's feedback on drafts can provide the student with material for text improvements (Eriksen, 2018). Students can also engage in a dialogue with both other students and generative AI as a conversation partner.

**Writing Argumentative Subject-Oriented Student Texts**

Writing a new text on a given topic is a challenge. It requires a thought process involving an analytical and systematic approach to find a solution to that challenge; it is a problem-solving process (Arnold et al., 2017). The concept of genre is important in our context. Here, we limit ourselves to argumentative factual writing (Berge, 2007; McCann, 1989), which involves developing reasoning using arguments. One step in the process of arguing a point is incorporating counterarguments, which is central in the scholastic genre of 'discussion'. Here, the text is often constructed around a central viewpoint which, through discursive argumentation, ends in a conclusion (Toulmin, 1958). A subject-oriented text can be improved if the student searches for counterarguments during the writing process (Bereiter, 2016). For many students, it can be mentally demanding to include arguments that go against their own views, and here AI can be used by assigning it the task of being a dissenting voice with a critical view of the student's text. When students are stuck, generative AI can help brainstorm ideas, plot points, or arguments, which can jump-start the writing process and lead to discovering new aspects of their voice. Including counterarguments can strengthen the text because a writer that shows openness to others' arguments and thinking can appear more credible to a reader – a point that can be tied to what rhetoric calls ethos appeal. A student who has developed the ability to see a case from multiple sides will be in a stronger position in many subjects. Study preparatory education programs are supposed to train students for higher education. Success at that level of learning requires a student to master the functional use of both written and oral language. Here too, the ability to present nuances that also includes counterpoints is a quality trait. Can AI be useful in this process?

**AI Integrated into a Writing Process**

Students who work with writing in stages tend to perform better than students who do not (Graham & Perin, 2007). Based on this insight, introductory books on writing instruction (e.g., Applebee & Langer, 2015) often recommend that writing takes place through various stages that are commonly divided into the pre-writing, writing, revision, and publishing phases.

Teachers and investigators are exploring how ChatGPT and similar AI-powered tools can transcend their reputation as facilitators of academic dishonesty and instead complement the array of aids students utilize in writing.

We now turn to how AI can be used in the various phases. Before students even begin, there are some factors that increase the chances of their text being good. One is identifying the audience. In schools, that is most often the teacher, but a writing assignment that can exceed this limitation and target another real recipient can increase the chances of students writing more effectively. The teacher can also assign tasks that have students write for authentic purposes, such as letters, blog posts, stories, or articles. This helps them see the value in developing a personal writing voice. Furthermore, students can choose topics about which they are passionate, which can contribute to a more engaged and authentic writing experience. Recipients could be class peers, younger students such as buddy children, the student council, the principal, a newspaper, or parents, but it may also be possible to instruct a chatbot to be the recipient and provide a considered response. We address this possibility further below.

In addition to having a clearly identified audience, the writing task should be clear and have appropriate criteria. This is also essential to increasing the chances that students will learn from an assessment, which is known as formative assessment or assessment for learning (Eriksen, 2018). The criteria should be few, as clear as possible, and be shared by both teacher and student to function to the best possible degree. In writing instruction, one should avoid general formulations like 'good structure' or 'correct spelling and punctuation' because they can be unclear or unattainable – what does it really mean to have good structure – and because writing correctly is something many people struggle with throughout their lives. It is better to focus on one thing at a time. For instance, every paragraph should have a topic sentence and discuss one thing – which will greatly contribute to good structure – or there should be particular attention to one type of error that many students make. This does not mean that we advocate for students not to strive to write correctly, just that it is difficult for a student to master everything at once, and that it is better to think that they are working on one thing at a time. Other examples of skills might be using certain technical terms, having arguments for and against a position, or making active use of academic sources, quoting them and commenting on those excerpts in their own presentation.

**Pre-writing Phase**

The main point in the pre-writing phase is to start the student's thought processes and to activate the resources they possess, both cognitively and socially. This idea is closely related to insights from reading research, where activating students' prior knowledge about a subject increases their chances of better understanding the text (Arnold et al., 2017). When students set about writing and developing their own voice in the process, it is important that they feel they have something substantial to write about. Here, it is common to use strategies such as brainstorming ideas, creating mind maps, and gathering points to be deployed in writing. When students have to write an argumentative text, finding arguments and counterarguments (that may involve a factual basis) will come into play (Gregg & Steinberg, 2016). They can also plan in a more targeted way how various rhetorical tactics like building credibility (ethos), appealing to emotions (pathos), and staying on topic and writing based on reasoning (logos) will be applied in their text (Berge, 2007). Perhaps one form of appeal at a time is enough when students are younger and still mastering the basics of the writing craft. Whether AI can be used as an idea generator in this phase is discussed below, but it is central to find a balance between having the student's own thoughts at the center and giving some help, a desired position, and the almost opposite situation, where AI provides the ideas, and the student does not have ownership of them.

Research on AI and its impact on creativity in this pre-writing phase presents a complex picture (Abdelghani, 2023). Dosher et al. (2024) show that generative AI tools like ChatGPT can enhance creative abilities in individuals who are otherwise considered less creative. However, these tools had little effect on already creative people. That research focuses on how AI tools can influence and possibly improve writing skills. The question of AI's role in the brainstorming phase requires a balance between writing support from AI and human imagination. AI may reduce cognitive load, but it can also threaten the development of depth in the pre-writing phase (Mollick & Mollick, 2023). It is possible that the collaboration between human and machine can enrich the creative process by bringing in new perspectives (Boynagryan & Tshngryan, 2024). However, it is not yet clear whether AI and human creativity are complementary or conflicting, and further research into and experience with this issue are necessary.

As part of the preparation for writing, or maybe even during the task, if students find themselves at an impasse (Gollins & Gentner, 2016), they can use generative AI as a devil's advocate when to find counterarguments to what they have written. This could be done, for example, by a Norwegian student entering the following prompt:

> I am very outraged that the political authorities of Norway do not stop the production of oil and gas. This is destroying our climate. We must stop destroying our climate. But the teacher says I need to discuss this and that I need counterarguments. Can you give me some counterarguments?

**Writing Phase**

In the writing phase, it is essential for students to use their own words and create a draft. It is not easy to see how AI can contribute sensibly in this phase without interfering with or even replacing the necessary process of the student formulating the text in their own language.

**Revision Phase**

Bereiter and Scardamalia's model for knowledge-transforming writing (2013) emphasizes that students must convert ideas into text in a way that not only represents what they know but also refines their understanding as they elaborate on the subject. The knowledge-transforming model involves a deeper level of cognitive engagement. Writers using that approach focus on conveying existing information and on reorganizing and reconceptualizing that information in the process. The model requires critical thinking, problem-solving, and the ability to make connections between ideas. The writing process here is more iterative and recursive, involving multiple drafts, revisions, and an ongoing process of refining thoughts and arguments. This sophisticated level of writing is often seen in more experienced writers and is associated with higher-order thinking and learning. Revising involves repeating certain steps or processes several times, often aiming to gradually improve the text. The student first creates a draft that provides a starting point for the improvement process. After this first draft has been finished, generative AI can be used to provide feedback (see below), based on which the student can then make changes to the draft. This can include rewriting parts of the text to increase clarity, reorganizing paragraphs to achieve better flow, or even discarding parts of the text that do not work as part of a whole.

**Feedback as Part of the Revision Phase**

Research on feedback on writing shows that it is particularly valuable for students to receive help while they are engaged in the writing task (Duijnhouwer et al., 2012; Wulandari, 2022). If students find themselves stuck or wonder how they have performed so far, feedback can be a great help (Eriksen, 2018). It is most likely to contribute to learning if it is concrete and contains explanations of what should be done to improve the text. Simply correcting errors is not enough. In addition, it is important that feedback points back towards the criteria defined for the task (Eriksen, 2018) and that students use it, where appropriate, to improve the text. Here, AI can be used to provide students with feedback. This can lighten the teacher's workload in the classroom, especially if there are as many as 30 (or more) students.

Unlike the human teacher, generative AI can provide instant feedback on a student's writing (Jeon, 2023). This might include highlighting strengths, suggesting improvements, and pointing out areas where the student's voice shines through (Jeon & Lee, 2024). It is crucial that AI receives clear instructions, such as checking whether there is a topic sentence that introduces each paragraph. One of the most important things here is that the prompt to AI be concrete and limited. Students can ask AI specific questions about their writing, such as 'What can I do to make my tone more professional?' This targeted feedback can guide them in refining their voice. Further, with the help of AI, students can receive feedback as soon as they have written, or they can pause and revise parts of the text while they are in progress and not feeling like they are finished. Providing feedback to everyone at this point in the process is beyond any teacher's reach; there simply is not enough time. In one example, a Swedish teacher (Haake & Gulz, 2024) shows how she allows the students to write a text without internet access; they then use AI to obtain feedback on their work, after which she disconnects the internet again while the students finish the assignment.

In the final writing phase, the focus is on finalizing the text for publication or submission. This is the point at which there is most to gain from working on technical errors of usage and grammar and general improvement of the language. The main reason for this is that many students, particularly those who struggle with writing, should not invest too much effort into the text too early and should rather orient themselves towards meaning production, ensure that their thoughts are clearly expressed, and that their responses meet the task criteria. By spending too much energy on correct language too early, students can lose sight of the overarching goals of the exercise. It can also be emotionally difficult to delete or remove paragraphs that do not respond to the task, simply because so much energy has already been invested in getting that language right.

An example of the first instruction to ChatGPT assigned the task of guiding in a relatively extensive writing process might look like the following. First, the teacher gives ChatGPT this prompt:

> This is a text written by a 10th grade student. Give two or three pieces of feedback that are specific, and that the student can use to improve their text. Focus on what the student has done well and what can be improved. Use a friendly and encouraging tone. If needed, provide the student with examples of how the text can be improved. (based on Steiss et al., 2024)

The student enters the following text:

I am a student in the 10th grade and have written an assignment. The goal is to improve the structure. The learner asks: "Can you suggest paragraph divisions and a topic sentence for each paragraph? Here is an authentic text":

> In this text, I will present facts and reflect on how the Norwegian authorities have influenced the Sami and other national minorities' language and culture, and the consequences of this. I will also look at the current situation and whether things have changed. Norway consists of many different types of people. One of them is the Sami. The Sami are an indigenous people with a traditional settlement from Norway, Sweden, Finland, and Russia. They have their own written and spoken language. The Sami also have a strong culture that is expressed through, for example, joik (a traditional form of song), kofte (traditional clothing), and reindeer herding. It is important for the Sami to showcase their culture, but have they always been able to do that?

In the 1800s, the Norwegification of the Sami began. This means that the Norwegian government worked to get the Sami to start speaking and writing in Norwegian instead of Sami. There were mainly two reasons for this. One of them was that Norway had just become an independent country and the Norwegians wanted everyone in the country to have the same language. The other reason was that the Norwegians did not want the Sami to have a separate religion.

Above, I have talked about why the Norwegians wanted to make the Sami Norwegian, but how were they supposed to do this? The Norwegification mainly took place in schools. This was done by having teaching in Norwegian. Many children lived away from their parents at boarding schools. At these schools, many experienced a lot of harm. It was common to hit students at that time, and if the children spoke Sami, they could be beaten. Norwegification also happened in other ways, including the government handing out Norwegian newspapers to the Sami. The Norwegification lasted a long time and in the mid-1900s, it was still practiced in many schools in Norway.

As I have said above, the Sami have been discriminated against for a long time through Norwegification and assimilation, that is, making the Sami Norwegian. When I think about how this has been, I understand that the Sami have lived in a society with a lot of racism. In addition, I think it must have felt difficult and hopeless for the parents of Sami children to send their children to schools where they knew the kids would be mistreated. The Sami are also a small group of people who I think did not have much to stand up against the Norwegian state. I also think that we have come a very long way both in school and in society in general; for example, I myself find it perfectly normal for people to speak languages other than Norwegian and use a language with a lot of multi-ethnic elements. A minority group that has been exposed to a lot of discrimination is the Kvens. The Kvens originate from Finnish immigrants and came to Norway in the 1700s. Like the Sami, they were also forced to attend boarding schools, where they were also subjected to abuse. Another form of discrimination is the fact that a law was passed in the Storting (the Norwegian parliament) in 1936 that made it forbidden to speak Kven in Norway. This led to the language almost being eradicated.

To summarize, I have talked about how the Sami and other minorities have been affected and discriminated against by the Norwegian state and provided examples of how they did this. In addition, I have reflected on how this might have been for those who experienced it and their parents.

AI responds as follows:

ChatGPT answers: Certainly! Here's how the text could be divided into paragraphs with a headline for each paragraph:

**Introduction to Sami Culture and Initial Impact:** In this text, I will present facts and reflect on the historical impact of Norwegian policies on the Sami and other national minorities. The Sami are an indigenous people with a unique culture that has faced significant challenges throughout history.

**The Norwegification Policy in the 1800s:** In the 1800s, the process of Norwegification represented the Norwegian government's concerted efforts to integrate the Sami into a singular Norwegian identity. This policy aimed to erase the linguistic and religious uniqueness of the Sami community in the name of national unity.

**Methods of Cultural Assimilation:** The methods of assimilating the Sami involved repressive educational strategies and other governmental actions. Mandatory schooling in Norwegian and the distribution of Norwegian literature were tactics used to suppress Sami language and culture.

**Reflections on Discrimination and Assimilation:** The long-term discrimination and enforced assimilation faced by the Sami can be regarded as a form of cultural erasure stemming from systemic racism. Considering the extensive discrimination endured, the resilience of the Sami and other minorities is noteworthy, especially when reflecting on the societal changes that have occurred since then.

**The Kven Minority and Broader Discrimination:** Discrimination against minority groups in Norway extends beyond the Sami, with the Kven experiencing similar injustices. The Kvens, descendants of Finnish immigrants, also faced severe oppression, including the banning of their language which led to its near extinction.

**Conclusion and Personal Reflection:** In conclusion, this text has outlined the Norwegian state's historical discrimination against the Sami and other minorities, illustrating the damaging policies and reflecting on their enduring effects. These reflections offer insights into the struggles of the indigenous Sami people and the broader societal implications of such policies.

This feedback is highly concrete, and the student can consider using the suggestions to improve the text while maintaining ownership of it. This shows that a short and highly specific request works better and aligns more closely with what research suggests is effective feedback than a longer and more open-ended prompt. Another reason for this is that a more open-ended prompt tends to produce more lengthy responses from the chatbot, which can be overwhelming especially for less proficient students.

In a study that compares the quality of feedback from generative AI and 'well-trained educators' (Steiss et al., 2024), the researchers found that humans provide better feedback. They used criteria from research about how feedback can best contribute to learning: it targets the task's criteria, includes an explanation of how the student should improve, is precise, focuses on the most important issues, and maintains a supportive tone. The researchers' findings illustrate how crucial it is for teachers to be closely involved with students and to assure the quality of the feedback and the related student work. This point is also highly relevant for self-assessment and peer assessment.

Coenen and coauthors (2021) have introduced Wordcraft, which is an AI-powered editing tool designed to facilitate story writing through a partnership between an author and a conversational AI system. Their interface leveraged the concept of few-shot learning alongside the inherent capabilities of conversational exchanges to accommodate diverse forms of interaction. This editing platform served as a creative space for authors to experiment with the limits of transformer-based language systems and lays the groundwork for possible integrations of human feedback into training processes and for the development of original assessment techniques.

Marzuki and coauthors (2023) explored the spectrum of AI writing aids and evaluated their effect on the writing skills of students, focusing on content and structure, from the perspective of teachers instructing English as a Second Language (ESL). Analysis of semi-structured interviews of four ESL educators from three separate Indonesian universities, highlighted the array of AI writing aids utilized in these teachers' instructional settings. There was a unanimous consensus among the educators on the positive impact of AI writing tools in refining their students' writing abilities, notably in terms of content quality and structural coherence.

**Use of AI as Part of Self-assessment and Peer Assessment**

The assessment researcher Madalena Taras (2010) emphasizes that the primary point of self-assessment is learning. She shows that both self-assessment and peer assessment can have beneficial effects on learning and points out some core guidelines for this type of work to function. Although her research was conducted before generative AI was introduced, it has clear transfer value. For students to get the most out of assessing their own performance, they first need training and practice in using criteria and linking them to their own texts. It is crucial, as noted above, that there be only a few criteria and that they be as specific as possible.

Practice and training are also core points in peer assessment work, and Taras (2017) and Topping (2009 & 2017) highlight that central success criteria are explicit goals – where the point of the activity is clearly communicated – and that these goals be translated into a few criteria, preferably developed in collaboration with students so that both students and teachers feel ownership over them. One way to concretize how criteria can look in practice is through the teacher's modelling – demonstrating how to work with a specific text and thinking aloud together with the class to evaluate suggestions on a student text provided by a chatbot. When students work independently or in pairs, there is also much to gain from the teacher moving around, offering help, and gaining an overview of the quality of the feedback students receive from both the chatbot and from one another.

Levine and co-authors (2024) examined the employment of ChatGPT by upper secondary students as an aid for crafting argumentative writings. Their inquiries focused on these questions: What instructions did students give to ChatGPT? And in what manner did they integrate the AI's responses into their work? The authors investigated screencasts of student interactions with ChatGPT and their peers during the stages of planning, composing, and evaluating their arguments, and their findings reveal that, in the stages of strategizing and composing, students predominantly solicited ChatGPT for conceptual inspirations, which they then expanded upon to formulate their individual stances. During the revision phase, their use of ChatGPT paralleled the application of Grammarly or comparable editing services. Comparisons of their own compositions with ChatGPT's output aided students in distinguishing their personal authorial styles and deepening their understanding of rhetorical strategies and their impact. The authors of this study suggest that ChatGPT can be assimilated into a collective, interactive writing framework, and students can engage with ChatGPT as a constructive writing asset without bypassing the essential stages of planning, drafting, and revising.

Participating in writing groups or workshops can provide a supportive environment for feedback and growth. However, generative AI can also act as a peer reviewer, offering constructive comments that

4push students to think critically about their writing choices and how they contribute to their distinctive voice. Engaging in a dialogue with generative AI can also help students practice and refine their voice in real time. The conversational format can make the writing process more dynamic and less intimidating. However, including peer review processes and collaborative projects can encourage students to engage more deeply with the material and rely less on AI-generated content.

**Cheating and AI-generated Texts**

Since the launch of ChatGPT in 2022, teachers have faced challenges in identifying and handling texts written by AI. Language arts teachers have also expressed frustrations over having to deal with AI-generated texts: 'There are increasingly more robot texts in schools. It's a disaster' (Osvold, 2023) and 'The school faces an existential threat. The benefit of using the technology is here and now: easier work and good grades. The price – poorly developed writing and thinking skills – will be paid later' (Wathne, 2023). While definitively proving that a student has copied text from a chatbot is challenging, AI-generated texts have certain distinct characteristics that the language arts teacher should note.

Several tools have been developed to reveal what is created by AI and what is not, but the accuracy and efficiency of these tools vary. The matter is also not easy for experts. Scarfe et al. (2024) conducted a rigorous study in which researchers submitted entirely AI-generated written assignments into the exam system in five different psychology modules at UK universities. They found that 94% of AI-generated submissions were not detected by these exam sensors. These sensors were experts. The grades given to AI submissions were significantly higher than those achieved by real students. As things currently stand, we cannot rely on tools designed to detect texts created by generative AI.

Fleckensten et al. (2024) examined teachers' ability to detect texts created by ChatGPT among students' work. In two studies, neither novice nor experienced teachers could reliably differentiate between AI-generated and student-written texts. Experienced teachers had a slightly better ability to make accurate assessments, but both groups were overly confident in their judgments. The effect of a text's real or perceived origin on quality assessment varied. The results of this study indicate that AI can produce texts that are difficult for teachers to recognize, representing a challenge for educational institutions in assessing student work. Moreover, AI tools are constantly evolving and improving – the total number of errors and hallucinations is decreasing, making it even harder to distinguish AI-generated from human-created text. Nevertheless, some linguistic traits stand out, which is explained in the next paragraph.

**Characteristics of ChatGPT Language**

Language models often follow specific patterns in text production, giving AI writing distinctive features (Shere, 2024). The text can be somewhat banal, and ChatGPT language often uses the third-person form. ChatGPT language can be repetitive, with a tendency to use certain verbs (for example 'address' or 'delve into') and end with a grandiose, comprehensive statement ('Success allows you to fulfil your ambitions and dreams, which can be incredibly satisfying and provide a deep sense of meaning in your life'). Texts produced by ChatGPT can also create fake sources and contain incorrect or entirely fictional facts and quotes. This creates challenges not only for tools designed to detect AI use but also for teachers and others who should and need to be able to find differences between human work and AI-generated materials.

A commonly used sentence starter is 'It is important to [note/understand/consider]'. AI-generated texts share similarities with lists of terms. They often contain numbered body paragraphs or multiple headings and subheadings. Paragraphs also commonly start with formulaic transition phrases. Here is an example of the first terms in each paragraph in an essay produced by ChatGPT: 'Firstly', 'In contrast', 'Moreover,' 'On the other hand', and 'To conclude'. AI-generated text tends to discuss how 'readers' are 'challenged' to 'confront' ideologies or are 'invited' to 'reflect' on central topics. ChatGPT also tends to exaggerate: Something can be a 'powerful symbol', while even everyday phenomena can be 'groundbreaking', 'important', 'esteemed', 'invaluable', 'indelible', 'essential', 'gripping', or 'profound'. The word 'successful' is used more often in American contexts than in Nordic contexts, and this indicates that ChatGPT is trained on a significant number of American texts. Writing can be referred to as 'weaving' a 'rich' or 'intricate tapestry', or the writer can 'paint' a 'vivid picture' (Steere, 2024).

**Implications for teachers' work**

What implications do students' shortcuts have for the teacher's work? Teachers can place greater emphasis on process-based assessments, where students' development and understanding are evaluated over time through drafts, discussions, and presentations, rather than relying solely on the final product (Hodges & Kirschner, 2024). Some educational portals make this possible. Schools and teachers may need to rethink how exam tasks are designed (Gobert et al., 2023& 2024). Creating tasks that require deep personal reflection, individual perspectives, or creative solutions is one way to make tasks less susceptible to using AI while promoting authentic student effort. Examples include 'Describe a turning point in your life and how it reflects a larger social movement or trend'. Teachers can also ask students to interview an older family member or a local resident and write a reflection based on that person's life story: 'What can your grandfather's experiences teach us about the values in our society today?' or 'Create a script for a podcast episode that explores how modernization impacts traditional crafts in your hometown' (Notebook LM is a useful device for making podcasts). Another example is requiring personal authenticity: 'How does Martin Luther King's "I Have a Dream" speech change your perception of freedom?' or 'If you had the chance to save a hundred strangers or one close friend, which would you choose and why?' These types of assignments encourage authenticity and are difficult for students to solve using AI because they require genuine human experience, emotions, and personal judgment. While some students may try to circumvent these requirements, this type of work often demands a form of qualitative assessment that challenges them to demonstrate their abilities in critical thinking and self-reflection.

It is possible to use writing exercises under controlled conditions (for example, a two-hour test without access to AI) and emphasize it when setting end-of-term grades. This is important for calculating learners. If this is a premise for grading, the incentive to use generative AI in homework is weakened. Schools and teachers may also need to rethink how exams are conducted to ensure that an exam reflects students' own skills and competencies. For example, there is software (for instance, Safe Exam Browser) that prevents students from leaving an examining without submitting their answer to the teacher, thus preventing the use of AI.

Few groups of subject teachers have as heavy a workload as language arts teachers; in secondary schools, they must currently cover several grades and they need adequate documentation for grading. Can technology ease the teacher's workload? Some believe so (Edwards et al., 2018; Jeon & Lee, 2024). Teachers can use generative AI to create personalized learning experiences for students. These models can analyze students' writing and responses and, at best, provide tailored feedback and suggest texts that align with the student's specific learning needs. This support can save teachers time and effort in creating personalized materials and feedback and allow them to focus on other aspects of teaching. Moreover, chatbots can respond immediately when students ask something (Jeon, 2024).

The most basic help can occur at a syntactic level; that is, identifying and correcting language errors. At a semantic level, large language models can be used to highlight grammatical inconsistencies and suggest strategies for improvement. The language models can also be used to identify opportunities for subject-specific improvement in language use and to produce summaries of difficult texts, which can help teachers highlight key points in a text in a way that aids students develop an in-depth understanding of the relevant content.

In a review of existing research, Jeon and Lee (2024) found that chatbots can provide some benefits comparable to those offered by human-human interaction in some domains. By contrast, chatbots were less effective than humans in areas that may require socially appropriate interpersonal elements, such as maintaining interactivity and providing sufficient information in elaborations. In most studies reviewed in Jeon and Lee (2024), students showed a higher degree of interactivity with human partners than with chatbots (e.g., Ockey & Chukharev-Hudilainen, 2021). These results may be related to the fact that students were aware that they were interacting with machines despite chatbots' human-like ability to have a 'conversation'. One possible implication for the teacher's work is that the decision to use chatbots in education should be guided by what teachers intend to achieve. For example, teachers aiming to evaluate and raise levels of grammar or vocabulary in students' utterances may consider using chatbots when human partners are unavailable.

On the other hand, human partners may be more effective as conversational partners than chatbots when tasks are intended to cultivate students' interaction competence or encourage them to provide sufficient information. While chatbots can be as effective as human partners for promoting short-term engagement and learning, their long-term use – especially if they reach the limits of reciprocity – may weaken students' interest in chatbot activities (Jeon, 2024).

Although there are assumptions about the influence that AI technology can have on education, there are still few studies examining how teachers can benefit concretely from it. Jeon's study (2024) sheds light on the interaction between ChatGPT and language teachers by specifically identifying how the parties can play complementary roles in the teaching process. Eleven language teachers were invited to integrate ChatGPT into their teaching over a two-week period. Afterwards, they shared their experiences through interviews and presented logs from their interactions with the technology. The analysis of the collected data revealed four key roles for ChatGPT – conversation partner, content provider, teaching assistant, and evaluator – and three important teacher roles – pedagogical decision-maker, guide for student exploration, and developer of students' ethical understanding related to AI.

The study results suggest that AI can be a valuable educational tool that enhances the teacher's work. By leading conversations, technology helps engage students and provide them with direct feedback. As a content provider, AI can enrich teaching by explaining complex topics or supplementing the teacher's effort with additional information. In the role of teaching assistant, technology eases the burden for teachers by handling standard questions and administrative tasks (Edwards et al., 2018). This frees up valuable time for teachers to focus on more important pedagogical tasks. Finally, AI in the evaluator role can contribute formative assessments and support students in realizing their academic strengths and weaknesses. Teachers are central to this partnership. They must balance different resources and high-quality pedagogical decisions, choosing the right moments to use AI and adapting its use to each student's needs and learning goals.

Regarding summative assessment, research suggests that the accuracy and consistency required for comparative assessment of student texts are not yet sufficient for AI to undertake such assessment tasks on its own (Pinot de Moira et al., 2022). How this will develop over time remains to be seen. It has, however, been suggested that teachers can use advanced language models to semi-automate the assessment of student work by identifying potential qualities and improvement areas in texts such as essays and other writing projects (Jeon, 2023). This can potentially free up a significant amount of time that can instead be used for individualized feedback to students. As noted above, the question of using this technology for plagiarism control remains open, which can help prevent cheating. Teachers' choices related to using AI tools should therefore be informed by the relationship between teacher and student, the individual student's personal needs, and the dynamics of the classroom.

Gee, J. P. (2012). What is literacy?. In: H. Luria, D.M. Seymour & T. Smoke (eds.) *Language and linguistics in context* (pp. 257-264). Routledge.

Genlott, A.A. & Grönlund, Å. (2013). Improving literacy skills through learning reading by writing: The iWTR method presented and tested. *Computers and Education*, 67, 98–104.

Gobert, J. D., Dickler, R., & Adair, A. (2024). Using an AI-Based Dashboard to Help Teachers Support Elevs' Learning Progressions for Science Practices. In: H. Jin, D. Yan & J. Krajcik (eds.) *Handbook of Research on Science Learning Progressions* (pp. 488-498). Routledge.

Gobert, J. D., Sao Pedro, M. A., & Betts, C. G. (2023). An AI-based teacher dashboard to support students' inquiry: Design principles, features, and technological specifications. In: J.D. Gobert, M.A. S. Pedro, C.G. Betts (eds.) *Handbook of Research on Science Education* (pp. 1011-1044). Routledge.

Gollins, A., & Gentner, D. (2016). A framework for a cognitive theory of writing. In L.W. Gregg & E.R. Steinberg (eds.) *Cognitive processes in writing* (pp. 51-72). Routledge.

Gregg, L. W., & Steinberg, E. R. (2016). *Cognitive processes in writing*. Routledge.

Haake, M. & Gulz, A. (2024). *Det oumbärliga klassrummet. En framställning på kognitionsvetenskaplig grund*. Natur & Kultur.

Hartwell, K., & Aull, L. (2023). Editorial Introduction–AI, corpora, and future directions for writing assessment. *Assessing Writing*, 57, 100769.

Hayes, J.R. & Flower, L.S. (1980). Identifying the Organization of Writing Processes. In: L.W. Gregg & E.R. Steinberg (eds.) *Cognitive Processes in Writing*. Routledge.

Hodges, C. B., & Kirschner, P. A. (2024). Innovation of instructional design and assessment in the age of generative artificial intelligence. *TechTrends*, 68(1), 195-199.

Hutson, J. (2024). Rethinking Plagiarism in the Era of Generative AI. *Journal of Intelligent Communication*, 4(1), 20-31.

Jeon, J. (2023). Chatbot-assisted dynamic assessment (CA-DA) for L2 vocabulary learning and diagnosis. *Computer Assisted Language Learning*, 36(7), 1338–1364. https://doi.org/10.1080/09588221.2021.1987272

Jeon, J. (2024). Exploring AI chatbot affordances in the EFL classroom: young learners' experiences and perspectives. *Computer Assisted Language Learning*, 37(1–2), 1–26. https://doi.org/10.1080/09588221.2021.2021241

Jeon, J. & Lee, S. (2024). Can learners benefit from chatbots instead of humans? A systematic review of human-chatbot comparison research in language education. *Education and Information Technologies*. https://doi.org/10.1007/s10639-024-12725-9

Jones, J. (2004). Learning to write in the disciplines: the application of systemic functional linguistic theory to the teaching and research of student writing. *Analysing academic writing*, 253-273.

Komakhidze, M., Reynolds, K.A., Fishbein, B., Wry, E., Kennedy, A., & von Davier, M. (2024). *COVID-19 Research Resources in PIRLS 2021*. Boston College, TIMSS & PIRLS International Study Center